\documentclass[%
 aip,
 amsmath,amssymb,
 reprint,%
]{revtex4-1}

\usepackage{graphicx}% Include figure files
\usepackage{dcolumn}% Align table columns on decimal point
\usepackage{bm}% bold math
\usepackage{epstopdf}

\usepackage[utf8]{inputenc}
\usepackage[T1]{fontenc}
\usepackage{mathptmx}
\usepackage{etoolbox}
\usepackage{float}
    \makeatletter
    \let\newfloat\newfloat@ltx
    \makeatother
\usepackage{algorithm}
\usepackage{algpseudocode}
\usepackage{xcolor}
\usepackage{hyperref} % hyperref must be loaded before apacite
\newcommand{\RNum}[1]{\uppercase\expandafter{\romannumeral #1\relax}}

\begin{document}

\preprint{AIP/123-QED}

\title[]
{Improved Memory Truncation Scheme for Quasi-Adiabatic Propagator Path Integral
via Influence Functional Renormalization}
%  {Soft Truncation of Memory Effect in Quasi-Adiabatic Propagator Path Integral for Open Quantum Dynamics}
\author{Limin Liu}
\author{Jiajun Ren}
\email{jjren@bnu.edu.cn}
\author{Weihai Fang}
\affiliation{Key Laboratory of Theoretical and Computational Photochemistry,
Ministry of Education, College of Chemistry, Beijing Normal University, 100875
Beijing, People’s Republic of China.}

\date{\today}% It is always \today, today,

\begin{abstract}
    Accurately simulating non-Markovian quantum dynamics in system-bath coupled
    problems remains challenging. In this work, we present a novel memory
    truncation scheme for the iterative Quasi-Adiabatic Propagator Path Integral
    (iQuAPI) method to improve accuracy. Conventional memory truncation in
    iQuAPI discards all influence functional beyond a certain time interval,
    which is not effective for problems with a long memory time. Our proposed
    scheme selectively retains the most significant parts of the influence
    functional using the density matrix renormalization group algorithm. We
    validate the effectiveness of our scheme through simulations of the
    spin-boson model across various parameter sets, demonstrating faster
    convergence and improved accuracy compared to the conventional scheme. Our
    findings suggest that the new memory truncation scheme significantly
    advances the capabilities of iQuAPI for problems with a long memory time.
\end{abstract}

\maketitle

\section{Introduction}
System-bath coupled models provide a microscopic description of dissipative
quantum processes in the condensed phase,~\cite{nitzan2006chemical} such as
spectroscopy and energy transfer in protein environments in light-harvesting
complexes,~\cite{ishizaki2009theoretical, yan2021theoretical} optoelectronic
processes in organic semiconductors,~\cite{wang2019hierarchical, li2021general}
and quantum transport in molecular electronic devices,~\cite{wang2011numerically,
jin2008exact} etc. The theoretical investigation of the dynamics of system-bath
models is essential for understanding the macroscopic phenomena observed in
experiments from a microscopic perspective. 

However, it remains a significant challenge to simulate the quantum dynamics of
system-bath models with high accuracy beyond the common perturbative and
Markovian approaches. The wavefunction approaches, including the multilayer
multiconfiguration time-dependent Hartree (ML-MCTDH)~\cite{wang2003multilayer, wang2009numerically}
and time-dependent density matrix renormalization group (TD-DMRG)
approaches,~\cite{ma2018time, ren2022time} treat the system and bath degrees of
freedom (DOFs) on an equal footing. For condensed phase problems, hundreds or
thousands of discrete bath modes need to be treated explicitly for a single bath,
making it difficult to extend to more complex problems with multiple baths. When
only the dynamics of the system part is of interest, it is more common to
calculate the reduced dynamics of the system by treating the bath DOFs
implicitly and a priori. Nevertheless, the memory effect induced by the bath
makes the reduced dynamics of the system non-Markovian --- the dynamics depends on
all the historical information of the system part, which is also called temporal
correlation.~\cite{breuer2016colloquium, sonner2021influence}

There are many numerically exact reduced quantum dynamics approaches that have
been proposed, including hierarchical equations of motion (HEOM),
~\cite{tanimura2020numerically, yan2004hierarchical, jin2008exact,
shi2009efficient} quasi-adiabatic propagator path integral (QuAPI),
~\cite{makri1992improved, makri1995tensor1, makri1995tensor2} stochastic
Schr\"{o}dinger equation,~\cite{strunz1999open, yan2016stochastic} generalized
quantum master equation (GQME),~\cite{mulvihill2021road, shi2003new} etc. Among them, QuAPI,
developed by Makri and coworkers, based on the Feynman-Vernon influence
functional theory,~\cite{feynman2000theory} can in principle handle arbitrary
bath spectral density and zero temperature, making it very compelling. The
practically used iterative QuAPI (iQuAPI) approach and its variants are not only
widely adopted to benchmark other methods but also applied to many real-world
problems.~\cite{kundu2020real,kundu2022tight, kundu2023pathsum} However, the computation and storage of the central quantity
in QuAPI, the augmented reduced density tensor (ARDT), increases exponentially
with respect to the physical memory time. To make numerical simulation feasible,
the memory time must be truncated in practice, known as the finite memory
approximation. The truncation length can be regarded as a parameter that needs
to be converged. For problems with long memory, it has been found that
convergence is very challenging with this conventional truncation scheme in
iQuAPI.~\cite{strathearn2017efficient}

Many algorithms have been proposed to improve the accuracy of QuAPI when dealing
with problems with long memory, including the path filtering,
~\cite{sim2001quantum} the blip decomposition,~\cite{makri2017iterative} the
coarse graining,~\cite{richter2017coarse} the scaling coefficient of influence
functional,~\cite{strathearn2017efficient}, the kink sum,~\cite{makri2024kink} etc.
The more recently developed time-evolving matrix product operator approach
(TEMPO) and its variants use matrix product operators to represent the influence
functional and matrix product states to approximate ARDT, which can greatly
improve the efficiency of QuAPI for problems with long
memory.~\cite{strathearn2018efficient, jorgensen2019exploiting,
ye2021constructing,bose2022multisite,ng2023real, bose2023quantum, link2024open, chen2024grassmann} The small
matrix decomposition of the path integral approach (SMatPI) and its extended
memory algorithm (x-SMatPI), developed by Makri, disentangles the original path
integral recursively and thus the residual terms are negligible and can be
discarded. As a result, the storage of the original exponentially large ARDT can
be replaced with small matrices with only two time indices.~\cite{makri2020small,
makri2020small2, makri2021small} 
%Although the effectiveness of TEMPO and SMatPI has been
%successfully demonstrated in some problems, more systematic benchmarks are necessary.

%%点题，指出本文的研究课题（我们工作是怎么做的，填补了怎样的空白or做了什么提高）我们方案的物理思想大致是什么，最终表现出来的效果是什么样的。
In this work, based on the iQuAPI approach, we propose a new memory truncation
scheme that improves the accuracy of the conventional scheme. The idea of the
scheme is that the influence functional beyond a certain time interval is not
fully discarded as in the original iQuAPI algorithm, but the most important part
is selected and retained according to the density matrix renormalization group
theory (DMRG).~\cite{white1992density, white1993density} In our method, an auxiliary time slice is introduced, which is the
renormalization of all the historical time slices beyond a certain time
interval. The effectiveness of our method is demonstrated by comparing it with
the conventional iQuAPI method for simulating quantum dynamics of the spin-boson
model with a relatively long memory time.

%%概括性介绍本文做了哪些工作（分别介绍文章的每一节都做了什么，有什么内容）
The remaining sections of this paper are arranged as follows: In Sec.~\ref{sec:
methods}, we briefly recap the iQuAPI algorithm and present the detailed
algorithm of our memory truncation scheme. In Sec.~\ref{sec:results}, we
show the numerical results of the quantum dynamics of the spin-boson model.
Finally, we present our conclusions in Sec.~\ref{sec:conclusion}.

%%%%%%%%%%%%%%%
%%Theory
%%%%%%%%%%%%%%%
\section{Methods}
\label{sec: methods}
\subsection{System-bath model and iQuAPI approach}

System-bath models are widely used to study quantum dissipative dynamics in the
condensed phase. In this work, we only consider bosonic baths. The generic
Hamiltonian is written as
\begin{align} 
    \hat H  & = \hat H_\textrm{S} + \hat H_\textrm{B} + \hat H_\textrm{SB}   \\
    \hat H_\textrm{B} & = \sum_i \frac{1}{2} \hat p_i^2 + \frac{1}{2} \omega_i^2 \hat q_i^2 \\
    \hat H_\textrm{SB} & =  \sum_{n} \hat{S}_n \otimes \sum_i c_{ni} \hat{q}_i \label{eq:H_SB}
\end{align}
Here $\hat H_\textrm{S}$ is the system Hamiltonian, which is assumed to be
simple to solve. $\hat H_\textrm{B}$ is the bath Hamiltonian, composed of
independent harmonic modes with frequency $\omega_i$ for mode $i$. $\hat
H_\textrm{SB}$ is the interaction between the system and bath, in which the system
operator $\hat S_n$ is linearly coupled to the coordinates of the bath with
coupling strength $c_{ni}$.

The reduced density matrix of the system part after tracing the whole density
matrix over the bath part is $\rho_\textrm{S}(t) = \textrm{Tr}_\textrm{B} \rho(t)
$. With reduced quantum dynamics approaches, an effective equation of motion of
$\rho_\textrm{S}(t)$ under the influence of the bath is to be solved.

The QuAPI method was proposed to solve the reduced quantum dynamics and has been
described in detail in a previously excellent review.~\cite{makri1998quantum} For
simplicity, we only consider one coupling term in Eq.~\eqref{eq:H_SB} and thus
the summation over $n$ is neglected. The extension to multiple coupling terms
with both diagonal and off-diagonal system-bath couplings is referred to
Ref.~\onlinecite{palm2018quasi,acharyya2020role,richter2022enhanced}. In QuAPI, the Hamiltonian is
re-partitioned into

\begin{gather}
    \hat H = \hat H_0 + \hat H_1 \\ 
    \hat H_0 = \sum_i \frac{1}{2} \hat p_i^2 +
        \frac{1}{2} \omega_i^2 (\hat q_i + \frac{c_i \hat{S}}{\omega_i^2})^2 \\ 
    \hat H_1 = \hat H_\textrm{S} - \sum_i \frac{c_i^2 \hat{S} ^2}{2\omega_i^2}
\end{gather}
With this partition, the formal propagator is split approximately by Trotter
decomposition. For clarity, the equations in this section are presented with
first-order Trotter splitting $e^{-i\hat H \Delta t} \approx e^{-i\hat H_0
\Delta t} e^{-i\hat H_1 \Delta t}$. However, in our calculations in
Sec.~\ref{sec:results}, the second-order Trotter splitting ($e^{-i\hat H \Delta
t} \approx e^{-i\hat H_1 \Delta t/2} e^{-i\hat H_0 \Delta t} e^{-i\hat H_1
\Delta t/2}$) is used for higher accuracy. The extension of the first-order
formulation to second-order formulation is straightforward, similar as
Ref.~\onlinecite{makri1995tensor1,makri1995tensor2}. After inserting
multiple resolutions of identity and integrating out the bath part analytically,
the evolution of the reduced density matrix of the system is expressed as the
path integral with $N$ time slices.

\begin{align}
\rho_S(N\Delta{t}) & = \textrm{Tr}_\textrm{B} \left \langle {s^+_N} \left |
    e^{-i\hat H N\Delta{t}}\rho(0) e^{i \hat H N\Delta{t}}\right |{s^-_N}\right
        \rangle  \nonumber \\
    & = \int  ds^+_0 \cdots \int  ds^+_{N-1} \int ds^-_0 \cdots \int ds^-_{N-1} \nonumber \\
    & \left \langle {s^+_0}\left | \rho_s(0)\right |s^-_0\right \rangle  B(s_0^\pm,
        s_1^\pm, \cdots, s_N^\pm) 
    F(s^\pm_1, s^\pm_2,\cdots,s^\pm_N) \label{eq:fvpi}
\end{align}
$\rho(0)$ is assumed to be factorized $\rho(0)=\rho_\textrm{S}(0)
\rho^{\textrm{eq}}_B$. The formulation to simulate the equilibrium initial state, please
refer to Ref.~\onlinecite{shao2001iterative, shao2002iterative}. $|s\rangle$ is the eigenstate of the system operator $\hat{S}$ with eigenvalue
$s$. The superscript $\pm$ indicates the forward and backward propagation,
respectively. For clarity, in the following the variable in parentheses of each
tensor is omitted if possible. $B$ is the bare system propagator written as 

\begin{align}
    B(s_0^\pm, s_1^\pm, \cdots, s_N^\pm) & = \prod_{k=1}^N K_{k-1,k} \\
    K_{k-1,k}  :&= K_{k-1,k}(s^\pm_{k-1}, s^{\pm}_{k}) \\ & = 
    \left \langle{s^+_{k}} \left |e^{-i\hat H_1\Delta{t}} \right |s^+_{k-1}\right \rangle
    \left \langle{s^-_{k-1}} \left |e^{i \hat H_1\Delta{t} } \right |s^-_{k}\right \rangle \label{eq:K}
\end{align}
$F$ is the discrete Feynman-Vernon influence functional, which can be decomposed
into the products of pairwise components $I_{k'k}$ ($k' \le k$). The influence
functional describes the influence of the bath on system dynamics and introduces
temporal correlation.

\begin{gather}
    F(s^\pm_1, s^\pm_2,\cdots,s^\pm_N) = \prod_{k=1}^{N} \prod_{k'=1}^{k} I_{k'k} \label{eq:IF} \\
    I_{k'k} := I_{k' k}(s^\pm_{k'},s^\pm_{k})= \exp \big[ -(s^+_k-s^-_k)(\eta_{k'k}s^+_{k'}-\eta^*_{k'k}
    s^-_{k'}) \big ] \label{eq:I}
\end{gather}
The coefficient $\eta_{k'k}$ only depends on $\Delta{k} = k-k'$, which is expressed as~\cite{makri1995tensor1}
\begin{equation}
\begin{aligned}
\eta_{k'k}&=\frac{2}{\pi }\int\limits_{-\infty}^{\infty}d\omega \frac{J(\omega)}
    {{\omega}^2}\frac{\exp(\beta \omega/2)}{\sinh(\beta \omega/2)}
    \sin^2(\omega{\Delta}t/2)e^{-i{\omega\Delta}t(k-k')}, \, k'<k  \\
\eta_{kk}&=\frac{1}{2\pi }\int\limits_{-\infty}^{\infty}d\omega \frac{J(\omega)}
    {{\omega}^2}\frac{\exp(\beta \omega/2)}{\sinh(\beta \omega/2)}
    (1-e^{-i{\omega\Delta}t}), \, k'=k 
\end{aligned}
\end{equation}
in which $J(\omega)$ is the bath spectral density, $J(\omega)= \frac{\pi}{2}\sum_i
\frac{c_i^2}{\omega_i}\delta(\omega-\omega_i)$. 

\begin{figure}
        \centering
        \includegraphics[width=0.5 \textwidth]{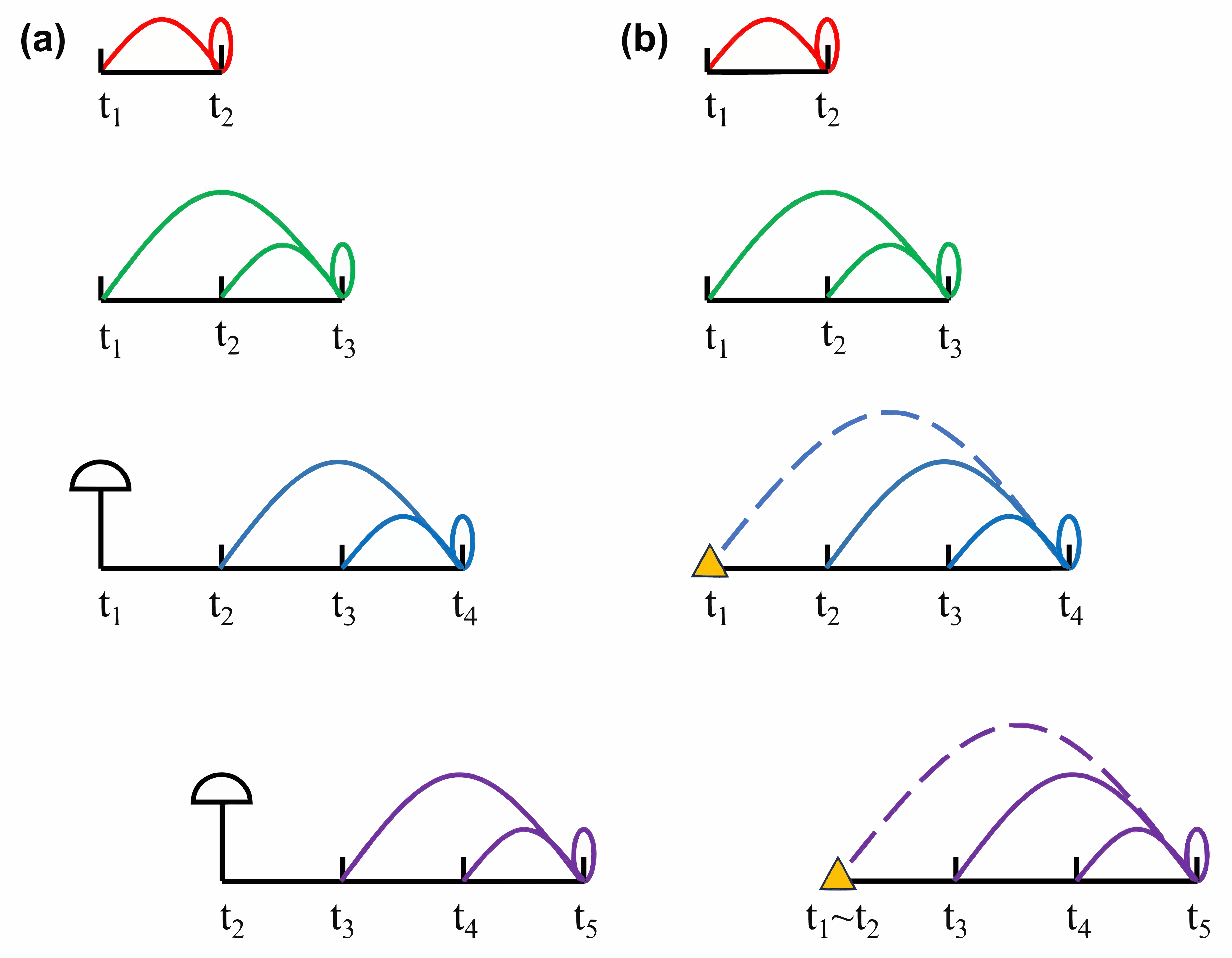}
        \caption{
            Schematic diagram of iQuAPI with (a) the conventional memory
            truncation scheme and (b) the improved memory truncation
            scheme based on DMRG. $\Delta k=2$. The lines in different colors
            represent the pairwise influence functional subterms $I_{k'k}$. A
            solid/dashed line indicates that the subterm is exact/approximated.
            The semicircle represents the summation of the time slices beyond
            the truncation length. The triangle represents the auxiliary time
            slice that contains partial information of the history.
            }
        \label{fig:fig1}
\end{figure}

To propagate Eq.~\eqref{eq:fvpi} in an iterative way, the influence functional
$F$ in Eq.~\eqref{eq:IF} can be grouped as

\begin{gather}
    F = \prod_{k=1}^{N} f_k \\ 
    f_k := f_k(s^\pm_1, s^\pm_2,\cdots,s^\pm_k) = \prod_{k'=1}^{k}
    I_{k'k} \label{eq:f_k}
\end{gather}
$f_k$ contains all the pairwise subterms of the influence functional between $k$
and $k'$ ($k' \le k$). 
With this decomposition of $F$, the key quantity in QuAPI called augmented
reduced density tensor $A_k$ can be iteratively calculated as
\begin{gather}
    A_k := A_{k}(s^\pm_1,\cdots,s^\pm_{k}) \\
    A_{k} = K_{k-1,k} f_{k} A_{k-1} \quad k>1 \label{eq:quapi} \\
     A_1 = \sum_{s_0^\pm} K_{01} f_1 \langle s_0^+ | \rho(0) |s_0^- \rangle
     \label{eq:A1}
\end{gather} 
The reduced density matrix of the system can be calculated from ARDT
\begin{equation}
    \rho_\textrm{S}(k\Delta{t}) = \sum_{s^\pm_1,\cdots,s^\pm_{k-1}}
    A_{k}. \label{eq:rho_quapi}
\end{equation}
The size of ARDT is exponentially increased with respect to the number of time
slices. Even for a two-state system, the affordable number of time slices is
less than 20. Fortunately, for a typical condensed phase problem, the physical
memory time $\tau_p$ is finite, which can be characterized by the bath
correlation time. More specifically, the pairwise
influence functional $I_{k'k}$ with $k-k' > \tau_p / \Delta t$ will become 1 as
$\eta_{k'k}$ approaches 0. In practice, we set a maximal time interval $\Delta
k$, when $k-k' > \Delta k$, $I_{k'k}$ is discarded. Accordingly, the influence
functional is truncated, that is
\begin{gather}
    f_k \approx \widetilde{f}_k := \widetilde f_k(s^\pm_{k-\Delta k}, \cdots,s^\pm_k) 
\end{gather}
With this truncation of influence functional, the time slices in ARDT before
$k-\Delta k$ will not be used anymore and thus can be integrated. The new
iterative equation when $k > \Delta k$ is
\begin{gather}
    \widetilde A_{k} := \widetilde A_k(s^\pm_{k-\Delta{k}+1},\cdots, s^\pm_{k}) \\
    \bar A_{k} := \bar A_k(s^\pm_{k-\Delta{k}},\cdots, s^\pm_{k})  \\
    \bar A_{k} = K_{k-1,k} \widetilde f_{k} \widetilde A_{k-1}, \quad 
     \widetilde A_{\Delta k} = A_{\Delta k} \label{eq:iquapi1} \\ 
    \widetilde A_{k} = \sum_{s_{k-\Delta{k}}^\pm} \bar A_{k} \label{eq:iquapi2} \\
    \rho_\textrm{S}(k\Delta{t}) = \sum_{s^\pm_{k-\Delta k+1},\cdots,s^\pm_{k-1}}
    \widetilde A_{k}.
\end{gather}
Following these iterative equations Eq.~\eqref{eq:iquapi1}\eqref{eq:iquapi2},
ARDT $\widetilde A_k$ will propagate with a fixed size $d^{\Delta k}$, where $d$
is the size of the system Hilbert space. This method is called iterative QuAPI.
Note that the intermediate tensor $\bar A$ does not have to be stored in the
actual implementation. Here, we keep it to make it easier to compare with the
new memory truncation scheme later. The schematic diagram of conventional iQuAPI
is shown in Fig.~\ref{fig:fig1}(a).

It should be emphasized that only when the preset maximal time interval is
longer than the physical memory time ($\Delta k \Delta t > \tau_p$), this
original memory truncation scheme in iQuAPI is exact. Otherwise, it is an
approximation, called finite memory approximation. Because the physical memory
time is not known in advance, in practice, several different $\Delta k$ should
be calculated to check the convergence. It has been known and will be shown in
Sec.~\ref{sec:results} that for problems with a pretty long physical memory time,
the results converge very slowly with $\Delta k$, and sometimes even cannot
converge in an affordable amount of computational cost. Therefore, we need a
better memory truncation scheme. This is the motivation of this work.

\subsection{Memory truncation scheme based on DMRG}
The main idea to improve the conventional truncation scheme is that the
influence functional beyond $\Delta k$ is not fully discarded, but is
renormalized and then the most important part of it is selected and retained.
The renormalization step is inspired by the density matrix renormalization group
theory~\cite{white1992density, white1993density} and accomplished by the
singular value decomposition (SVD) algorithm.

When the time step $k = \Delta k + 1$ (the first truncation step is a bit
different from the following steps), instead of summing $\bar{A}_{\Delta k + 1}$
over $s_1^\pm$ as in conventional iQuAPI (Eq.~\eqref{eq:iquapi2}), $\bar{A}
_{\Delta k + 1}$ after unfolding is decomposed by SVD as
\begin{align}
    \bar A_{\Delta k+1} \to \bar A_{\Delta k+1}(s^\pm_1|s^\pm_2 \cdots
    s^\pm_{\Delta k+1}) 
    & = \sum_{r=1}^{n} U_{s^\pm_1,r}\Lambda_{r} V^\dagger_{r, s^\pm_2
    \cdots s^\pm_{\Delta k+1}} \nonumber \\
    & \approx \sum_{r_1=1}^{M_1} U_{s^\pm_1,r_1}\Lambda_{r_1} V^\dagger_{r_1,
    s^\pm_2 \cdots s^\pm_{\Delta k+1}} \label{eq:svd0}
\end{align}
where $U$ and $V$ are column-wise orthonormal matrices. $\Lambda$ is a diagonal
matrix with real and non-negative diagonal elements ($\Lambda_{1} \ge \Lambda_{2}
\ge \cdots \ge \Lambda_{n} \ge 0$), called singular values. In the first
truncation step, $n=d^2$. Instead of keeping all the $n$ components, we set a
cutoff $\xi$ and select only the components with $\Lambda_{r}/ \sqrt{\sum_{r}
\Lambda_{r}^2} > \xi$, the size of which is $M_1$.  
This selection minimizes the Euclidean distance between the original ARDT and the
approximated ARDT. After this selection, the state $| s^\pm_{1} \rangle$ is
renormalized to $| r_1 \rangle$ according to $U_{s_1^\pm, r_1}$. 
\begin{gather}
   |r_1\rangle  = \sum_{s_1^\pm} |s_1^\pm\rangle U_{s_1^\pm,r_1} 
\end{gather}
$|r_1\rangle$ can be regarded as the basis states of the auxiliary time slice.
The renormalized ARDT is defined as 
\begin{gather}
    \widetilde A_{\Delta k+1}(r_1,s_2^\pm,\cdots,s_{\Delta k+1}^\pm) = \Lambda_{r_1} V^\dagger_{r_1,
    s^\pm_2 \cdots s^\pm_{\Delta k+1}}
\end{gather}
In addition, the pairwise influence functional between $s_1^\pm$ and $k''$ ($k''>
\Delta k+1$) is renormalized as
\begin{gather}
    \widetilde{I}_{1k''} (r_1,r_1',s^\pm_{k''}) = \sum_{s^\pm_1} U^\dagger_{r_1, s^\pm_1}
    I_{1k''}(s^\pm_1,s^\pm_{k''}) U_{s^\pm_1,r_1'} 
\end{gather}
$\widetilde{I}_{1k''} (r_1,r_1',s^\pm_{k''})$ is the influence functional
between the auxiliary time slice and normal time slice $k''$, which is a
three-legged tensor different from the original one. 
Correspondingly, 
\begin{gather}
    \widetilde{f}_{k''} (r_1,r_1',s^\pm_2, \cdots, s^\pm_{k''}) = \widetilde{I}_{1k''} 
      \prod_{k'=2}^{k''} I_{k'k''}
\end{gather}

To calculate the reduced density matrix of system at this time, it should be
noted that the renormalized states $|r_1\rangle$ have a different weight to the
original $| s_1^\pm \rangle$. The weight is 
\begin{gather}
    W_1(r_1) = \sum_{s_1^\pm}  U_{s^\pm_1,r_1} 
\end{gather}
therefore, 
\begin{gather}
\rho\big((\Delta k+1)\Delta{t}\big)=\sum_{r_1,s^\pm_2,\cdots,s^\pm_{\Delta k}} W_1(r_1)
    \widetilde A_{\Delta k+1}(r_1,s^\pm_2,\cdots,s^\pm_{\Delta k+1})
\end{gather}
It can be checked that when $r_1$ is not truncated ($\xi=0, M_1=d^2$), $\rho\big((k+1)
\Delta t\big)$ is exact, because only a unitary basis rotation between $|s_1^\pm
\rangle$ and $|r_1 \rangle$ is performed which does not alter the results. (See
Supplementary Material (SM) for the proof.) 

The renormalization step will continue iteratively. When $k > \Delta k+1$,
\begin{gather}
    \widetilde A_{k} := \widetilde A_{k} (r_{k-\Delta k},s^\pm_{k-\Delta k+1}, \cdots,s^\pm_{k}) \\
    \bar A_{k} := \bar A_{k} (r_{k-\Delta k-1},s^\pm_{k-\Delta k},\cdots,
    s^\pm_{k}) \\
    \bar A_{k} = \sum_{r'_{k-\Delta k-1}} K_{k-1,k}
        \widetilde{f}_{k} \widetilde{A}_{k-1} (r'_{k-\Delta k-1},s^\pm_{k-\Delta
        k}, \cdots,s^\pm_{k-1}) \label{eq:st-Abar}
\end{gather}
Similar as Eq.~\eqref{eq:svd0},
\begin{align}
    \bar A_{k}  & \to   \bar A_{k} (r_{k-\Delta k-1} s^\pm_{k-\Delta k} |
    s^\pm_{k-\Delta k+1} \cdots s^\pm_{k})
    \nonumber \\ 
    & \approx  \sum_{r_{k-\Delta k}=1}^{M_{k-\Delta k}} U_{r_{k-\Delta k-1}
    s^\pm_{k-\Delta k},r_{k-\Delta k}} \Lambda_{r_{k-\Delta k}}
    V^\dagger_{r_{k-\Delta k}, s^\pm_{k-\Delta k+1} \cdots s^\pm_{k}}
    \label{eq:svd} \\
    \widetilde A_{k} & =  \Lambda_{r_{k-\Delta k}} V^\dagger_{r_{k-\Delta k},
    s^\pm_{k-\Delta k+1} \cdots s^\pm_{k}} \label{eq:st-Atilde}
\end{align}
The state $|r_{k-\Delta k-1} \otimes s^\pm_{k-\Delta k} \rangle$ is renormalized to
$|r_{k-\Delta k}\rangle$ according to $U_{r_{k-\Delta k-1} s^\pm_{k-\Delta k},
r_{k-\Delta k}}$,
\begin{gather}
    |r_{k-\Delta k}\rangle = \sum_{r_{k-\Delta k-1}, s^\pm_{k-\Delta k}}
    |r_{k-\Delta k-1} s^\pm_{k-\Delta k} \rangle U_{r_{k-\Delta k-1}
    s^\pm_{k-\Delta k},r_{k-\Delta k}}
\end{gather}
The pairwise influence functional between $r_{k-\Delta k-1}$ and
$k''$ ($k''> k$), together with $ s^\pm_{k-\Delta k} $ and $k''$, is
renormalized to

\begin{gather}
    \widetilde{I}_{k-\Delta k, k''}:=\widetilde{I}_{k-\Delta k, k''}(r_{k-\Delta
    k},r_{k-\Delta k}',s^\pm_{k''}) \\
    \widetilde{I}_{k-\Delta k, k''} 
    = \sum_{r_{k-\Delta k-1},r_{k-\Delta k-1}', s^\pm_{k-\Delta k}} U^\dagger_{r_{k-\Delta k}, r_{k-\Delta k -1}
    s^\pm_{k-\Delta k}} \times  \nonumber \\ \widetilde I_{k-\Delta k-1,k''}
    I_{k-\Delta k, k''} U_{r'_{k-\Delta k -1}
    s^\pm_{k-\Delta k},r'_{k-\Delta k}} \label{eq:st-Itilde}
\end{gather}
Correspondingly,
\begin{gather}
    \widetilde{f}_{k+1} = \widetilde{I}_{k-\Delta k, k+1} \prod_{k'=k-\Delta k+1}
    ^{k+1} I_{k',k+1} \label{eq:st-ftilde}
\end{gather}
The weight of the renormalized states $|r_{k-\Delta k}\rangle$ is
\begin{gather}
    W_{k-\Delta k}:= W_{k-\Delta k}(r_{k-\Delta k}) \\
    W_{k-\Delta k} = \sum_{r_{k-\Delta k -1}, s^\pm_{k-\Delta
    k}}W_{k-\Delta k-1}  U_{r_{k-\Delta k -1} s^\pm_{k-\Delta k},r_{k-\Delta k}}
    \label{eq:st-w}
\end{gather}
The reduced density matrix of system is 
\begin{gather}
\rho(k\Delta{t})=\sum_{r_{k-\Delta k},s^\pm_{k-\Delta k+1},\cdots,
    s^\pm_{k-1}} W_{k-\Delta k}\widetilde A_{k} \label{eq:st-rho}
\end{gather}
The overall algorithm is shown in \textbf{Algorithm}~\ref{algo:algo1} and the schematic
diagram is shown in Fig.~\ref{fig:fig1}(b).

\begin{algorithm}
\caption{iQuAPI algorithm with memory truncation scheme based on DMRG}
\label{algo:algo1}
\begin{algorithmic}[1]
    \Procedure {prepare system propagator } {$\hat H_1$, $\Delta{t}$} 
        \State $K_{k-1,k} \gets \hat H_1, \Delta t$
        \Comment{Eq.\eqref{eq:K}} 
        \State \textrm{return} $K_{k-1,k}$
    \EndProcedure
    \Statex
    
    \Procedure{prepare influence functional }{$\eta_{k'k}$}
        \For{$k=1 \to N$}
        \For{$k'=1 \to k$}
            \State $I_{k'k} \gets \eta_{k'k}$ \Comment{Eq.\eqref{eq:I}}
            \State $f_k \gets I_{k'k}$ \Comment{Eq.\eqref{eq:f_k}}
        \EndFor
        \EndFor
        \State \textrm{return} $I_{k'k}$, $f_k$
    \EndProcedure
    \Statex

    \Procedure{evolution process } {$I_{k'k}$, $f_k$, $K_{k-1,k}$}
    \State $A_1,\rho(\Delta t) \gets K_{01},
        f_1, \rho(0)$ \Comment{Eq.\eqref{eq:A1}}
    \For{$k = 2 \to N$}
    \If{$k \le \Delta{k}$}
        \State $A_k \gets A_{k-1},f_k, K_{k-1,k}$  \Comment{Eq.\eqref{eq:quapi}} 
        \State $\rho(k\Delta t) \gets A_k$ \Comment{Eq.\eqref{eq:rho_quapi}}
        \State \textrm{if} $k=\Delta k$, $\tilde A_{\Delta k}, \tilde{f}_{\Delta k +1} \gets  A_{\Delta k},
        f_{\Delta k +1}$
    \Else
        %\If{$k=\Delta{k}+2$}
        %    \State $U,S,V^\dagger \gets \textbf{SVD} \left[ A_{k-1}(s^\pm_1|s^\pm_2 \cdots
        %    s^\pm_{k-1}) \right]$ \Comment{Eq.\eqref{eq:}}
        %\Else
        %    \State $U,S,V^\dagger \gets \textbf{SVD} \left[ A_{k-1}(r_{k-\Delta{k}-2},
        %    s^\pm_{k-\Delta{k}-1}| s^\pm_{k-\Delta{k}} \cdots s^\pm_{k-1}) \right]$ \Comment{Eq.\eqref{eq:}}
        %\EndIf
        \State $\bar A_k \gets \tilde{A}_{k-1}, \tilde{f}_k, K_{k-1,k} $
        \Comment{Eq.\eqref{eq:st-Abar}}
        \State $ U,\Lambda,V^\dagger \gets \textbf{SVD} \left[ \bar A_{k} \right]
        $ \Comment{Eq.\eqref{eq:svd}}
        \State $\tilde{A}_{k} \gets \Lambda, V^\dagger $ \Comment{Eq.\eqref{eq:st-Atilde}}
        \State $\tilde{I}_{k-\Delta k,k''} \gets U, \tilde I_{k-\Delta k-1, k''},
        I_{k-\Delta k, k''}$ \Comment{Eq.\eqref{eq:st-Itilde}}
        \State $\tilde{f}_{k+1} \gets \tilde I_{k-\Delta k,k+1}, I_{k',k+1}$ \Comment{Eq.\eqref{eq:st-ftilde}}
        \State $W_{k-\Delta k} \gets U, W_{k-\Delta k-1}$ \Comment{Eq.\eqref{eq:st-w}}
        \State $\rho(k \Delta t) \gets W_{k-\Delta k}, \tilde{A}_k$
        \Comment{Eq.\eqref{eq:st-rho}}
    \EndIf
    \EndFor
    \State \textrm{return} $\rho(k \Delta t)$
    \EndProcedure
\end{algorithmic}
\end{algorithm}

Compared to the memory truncation scheme in the original iQuAPI, where the
influence functional is sharply truncated when the time interval exceeds a
preset $\Delta k \Delta t$, the new truncation scheme introduces an auxiliary
time slice. The renormalized pairwise influence functional between the auxiliary
time slice and the future time slices retains the most important memory effect
beyond the preset maximal time interval, though it is not exact.  As a result,
we expect that the new memory truncation scheme can improve the results of
conventional iQuAPI, especially for problems with long memory.  Due to the
different features of these two truncation schemes, we refer to the new
truncation scheme as the soft truncation scheme (ST) and the conventional
truncation scheme as the hard truncation scheme (HT) in the following sections.

In ST-QuAPI, the basis states of the auxiliary time slice are renormalized
recursively and selected adaptively during the time propagation according to the
singular values $\Lambda_r$ of ARDT and the preset cutoff $\xi$.  Unlike
HT-QuAPI in which only the parameter $\Delta k$ determines the accuracy, in
ST-QuAPI, the SVD cutoff $\xi$ also determines the accuracy. 
When the cutoff $\xi$ is 0, meaning that all the basis states are retained, the
approach reverts to the exact QuAPI no matter what $\Delta k$ is, yielding exact
results since any unitary transformation between the basis states does not alter
the results.
Otherwise, the distribution of the singular values $\Lambda_r$ with the index of
renormalized state $r$ is very important, which characterizes the strength of
temporal correlation. When $\Lambda_r$ decays very fast with
$r$, only a small number of renormalized states need to be retained, in this
case the new truncation scheme is effective.  When $\Lambda_r$ decays very
slowly with $r$, a large number of renormalized states should be retained to
ensure a high accuracy. The worst case is that $\Lambda_r$ is equal for
different $r$, indicating that the temporal correlation between the historical
time slices and future time slices at this time point is extremely strong. In
this case, all the states must be retained and any truncation scheme is invalid.

We briefly analyze the computational scaling of the ST-QuAPI method.
For HT-QuAPI, the computational scaling is $O(d^{2\Delta{k}})$ from the
contraction of $A_{k-1} f_k K_{k-1, k}$. Compared to HT-QuAPI, the
contraction in ST-QuAPI scales as $O(M^2 d^{2\Delta{k}})$ because of the extra
auxiliary time slice. In addition, an SVD process is needed to obtain the
renormalized basis states in each time step, whose scaling is also $O(M^2
d^{2\Delta{k}})$. Thus, the overall computational scaling of ST-QuAPI is $O(M^2
d^{2\Delta{k}})$. In addition to the computational cost, the memory requirement to store the
high dimensional tensor in ST-QuAPI is $O(M d^{2\Delta{k}})$, while that in
HT-QuAPI is $O(d^{2\Delta k})$. Although the formal scaling of ST-QuAPI seems
larger than that of HT-QuAPI, the size of $\Delta k$ is much smaller for
ST-QuAPI than HT-QuAPI to get converged results.

Before closing this section, we will briefly discuss the relation between our
memory truncation scheme and TEMPO. As mentioned in the introduction, TEMPO
takes advantage of matrix product states (MPS) to approximate ARDT thereby preventing
ARDT from growing exponentially with time steps. Similarly, the iterative
renormalization step in our memory truncation scheme also essentially constructs
an MPS. 
The main difference lies in that in our approach the renormalization matrices
are fixed beyond the truncation length, while the matrices of MPS in TEMPO are
all allowed to be optimized. This difference is similar to that between infinite
DMRG and finite DMRG.~\cite{schollwock2011density} Although allowing the
matrices to be optimized makes TEMPO
more accurate, it is more expensive for long-time dynamics. Therefore, in
practical calculations by TEMPO, the conventional memory truncation scheme is
still being used.~\cite{strathearn2018efficient} In this regard, our new memory truncation scheme is compatible
with TEMPO to replace the currently used scheme to make it more accurate to
capture the memory effect. In other words, we can use TEMPO to
approximate the exact part of ARDT in ST-QuAPI to enlarge the parts where the
influence functional is treated exactly.

\begin{figure*}[htbp]
        \centering
        \includegraphics[width= \textwidth]{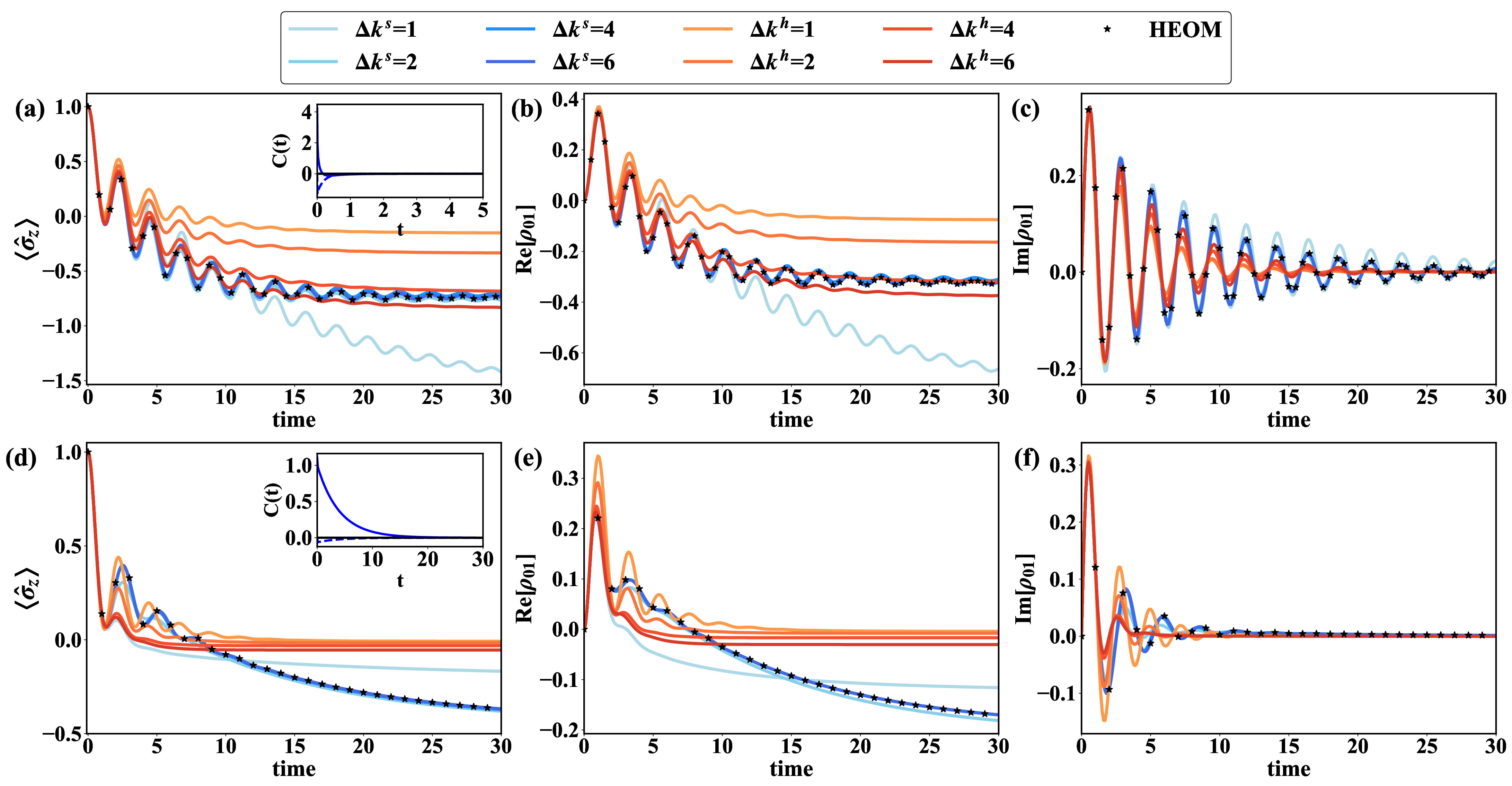}
        \caption{
            The reduced density matrix $\rho_\textrm{S}(t)$ of SBM with Debye
            spectral density. (a-c) $\langle \hat \sigma_z(t)\rangle$, $\textrm{Re}
            [\rho_{01}(t)]$ and $\textrm{Im} [\rho_{01}(t)]$ of SBM with $\beta=50,
            \omega_c=5$ calculated by ST-QuAPI (blue lines) and HT-QuAPI
            (conventional iQuAPI, red lines) with different truncation length
            $\Delta k=1, 2, 4, 6$. The black asterisk is the reference
            calculated by HEOM. The inset is the bath time correlation function
            (The solid line is the real part and the dashed line is the imaginary part).
            (d-f) same as (a-c) but with parameter $\beta=0.5, \omega_c=0.25$.
        } 
        \label{fig:fig2}
\end{figure*}

\section{Results}
\label{sec:results}

In this section, we will benchmark the new truncation scheme in the spin-boson
model (SBM). SBM is one of the most celebrated models to study quantum
dissipative dynamics in the condensed phase~\cite{nitzan2006chemical,
leggett1987dynamics}. It serves as a testbed to benchmark different quantum
dynamics methods. The Hamiltonian of SBM is written as

\begin{gather}
    \hat H= \varepsilon \hat \sigma _z + \Delta \hat \sigma_x
    + \sum_{i} \frac{1}{2}(\hat p_i^2 + \omega_i^2\hat x_i^2) + \hat \sigma_z \sum_{i} c_i \hat x_i
\end{gather}
We consider four cases with different parameters.

\subsection{SBM with Debye spectral density}
The first two cases adopt the Debye spectral density, $J(\omega)= \eta
\frac{\omega \omega_c}{\omega^2 + \omega_c^2}$. $\Delta=1$ is used as the unit. Case
\RNum{1} has a low temperature $\beta=50$ and fast bath motion $\omega_c=5$; Case
\RNum{2} has a high temperature $\beta=0.5$ and slow bath motion $\omega_c=0.25$.
The other parameters are $\varepsilon=1$, $\eta=0.5$.~\cite{song2016alternative, gao2022non} We simulate the spin dynamics by iQuAPI with the proposed soft memory
truncation scheme and the conventional hard truncation scheme. The reference
results are calculated with HEOM by QuTiP~\cite{JOHANSSON20131234} with $K=70,
L=3$ for Case \RNum{1} using Padé expansion and $K=4, L=20$ for Case \RNum{2} using Matsubara expansion. 

Fig.~\ref{fig:fig2}(a)(c) show $\langle \hat \sigma_z(t) \rangle$ of these
two cases with different $\Delta k$. The truncation cutoff of ST is $\xi=10^{-5}$.
The time step size is 0.1. The insets are the bath time
correlation function of the two cases characterizing the length of memory time.
\begin{gather}
    C(t)= \frac{1}{\pi}\int_0^{\infty} d \omega J(\omega)
    \left[\operatorname{coth} \frac{\beta \omega}{2} \cos \omega t-i \sin \omega
    t\right]
\end{gather}
Case \RNum{2} with slow bath motion has a relatively long memory time (> 10), while
Case \RNum{1} with fast bath motion has a short memory time (< 1). 
For the short memory case, ST (blue line) converges much faster than HT (red
line) with respect to $\Delta k$. When $\Delta k=6$, ST has achieved an accuracy
where the maximal absolute error $\varepsilon_m$ is smaller than $0.01$ within
$t \le 30$. For comparison, when $\Delta k=6$, even though the curve of HT is
qualitatively correct, $\varepsilon_m$ is about $0.1$. Even when $\Delta k=13$,
$\varepsilon_m$ is $\sim 0.03$. 
For the long memory case, the improvement of ST is much more pronounced. With HT,
the convergence of $\langle \hat \sigma_z(t) \rangle$ with $\Delta k$ is very slow.
In addition, the asymptotic behavior with $t$ seems totally wrong. On the
contrary, the error of ST is greatly reduced even with a very small $\Delta k$.
With $\Delta k=6$, $\varepsilon_m$ has already been smaller than $0.02$.
Besides the diagonal elements of the reduced density matrix, the off-diagonal
matrix elements are shown in Fig.~\ref{fig:fig2}(b),(c),(e),(f). The off-diagonal matrix
elements characterize the coherence between the two system states, and are
believed to be more difficult to calculate accurately than the diagonal
elements. In both of these cases, ST can obtain coherence much more accurately
than HT with the same $\Delta k$. Compared to HT, the improvement in both
population and coherence demonstrates that the most important memory effect is
indeed captured by the renormalization process in ST as expected. This
improvement is essential for problems with long memory time as Case
\RNum{2} shown in Fig.~\ref{fig:fig2}(d-f).

\begin{figure}[htbp]
        \centering
        \includegraphics[width=0.39 \textwidth]{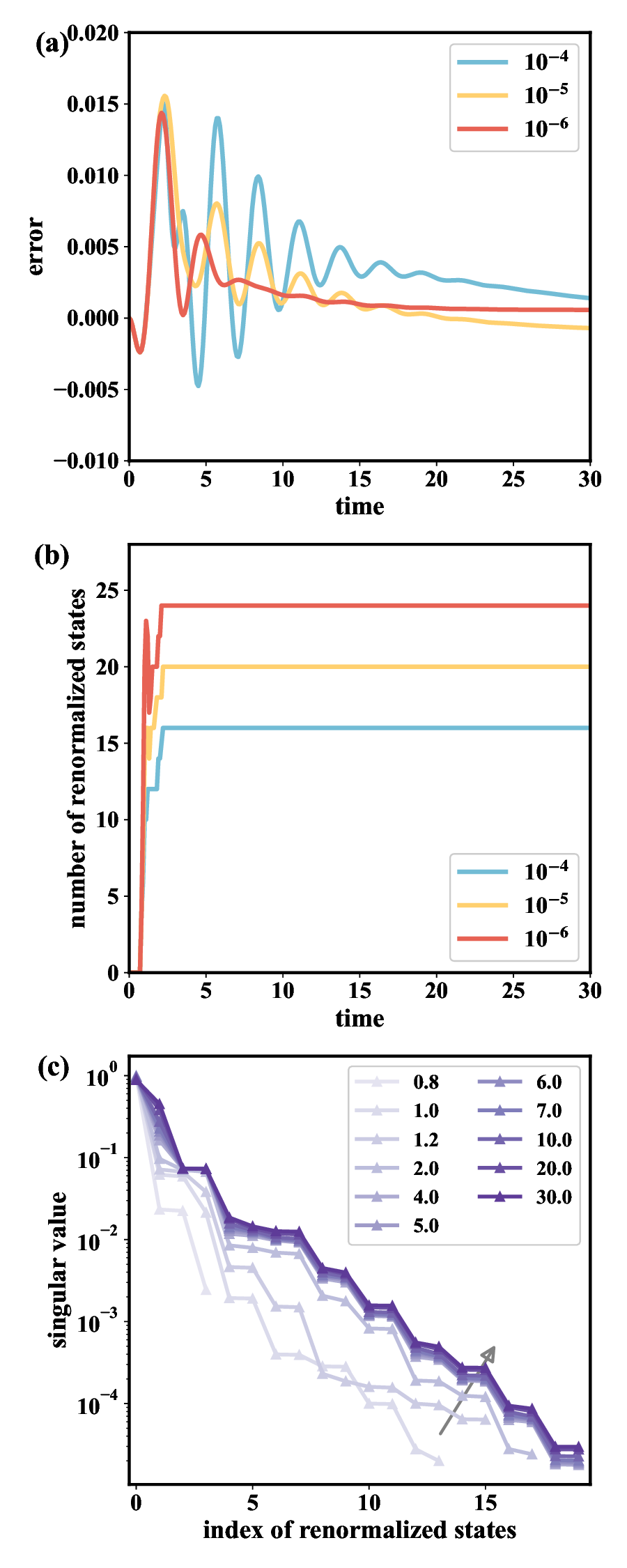}
        \caption{
            (a) The error of $\langle \hat \sigma_z(t) \rangle$ and (b) the
            number of renormalized states of the auxiliary time slice with SVD
            cutoff $\xi = 10^{-4},10^{-5},10^{-6}$ for Case \RNum{2}. (c) The
            distribution of singular values $\Lambda_r$ after normalization at
            different times from $t=0.8$ to $t=30$ with $\xi=10^{-5}$.}
        \label{fig:fig3}
\end{figure}

Besides $\Delta k$, the cutoff $\xi$ in SVD is also important for the accuracy
of ST-QuAPI. To evaluate the error brought by SVD cutoffs, we set a series of
cutoffs with the same $\Delta{k} =6$ for the long memory case, Case \RNum{2}. In
Fig.~\ref{fig:fig3} (a), we will get more accurate results if we take a smaller
cutoff. With $\xi < 10^{-4}$, the maximal absolute error is less than 0.02. In
Fig.~\ref{fig:fig3} (b), the number of renormalized states of the auxiliary time
slice grows rapidly at the beginning (during the time comparable to the bath
correlation time), and then it reaches a certain value which will become larger
as the cutoff $\xi$ decreases. The final number of renormalized states is
$15 \sim 25$. 

To monitor the decay of singular values $\Lambda_r$, which indicates the
strength of temporal correlation, we show the singular values at several time
steps in Fig.~\ref{fig:fig3}(c). Overall, with the parameters we considered, the
singular values decay quickly with the index of renormalized states. Moreover,
although the singular values increase with time suggesting the increase of
temporal correlation, they gradually converge to a certain value, meaning that
the temporal correlation is bounded with time.  This is
consistent with what is observed in the number of kept renormalized states with
a fixed cutoff in Fig.~\ref{fig:fig3}(b). The two observations in Fig.~\ref{fig:fig3}(c) that the singular values decay fast and the temporal correlation is
bounded, ensure the effectiveness of the new truncation scheme based on DMRG.
The data for Case \RNum{1} is similar to Case \RNum{2} (See SM Fig.~S1).  However,
whether the observation is general for other more challenging parameter regimes
needs further investigation.

\begin{figure*}[htbp]
    \centering
    \includegraphics[width= \textwidth]{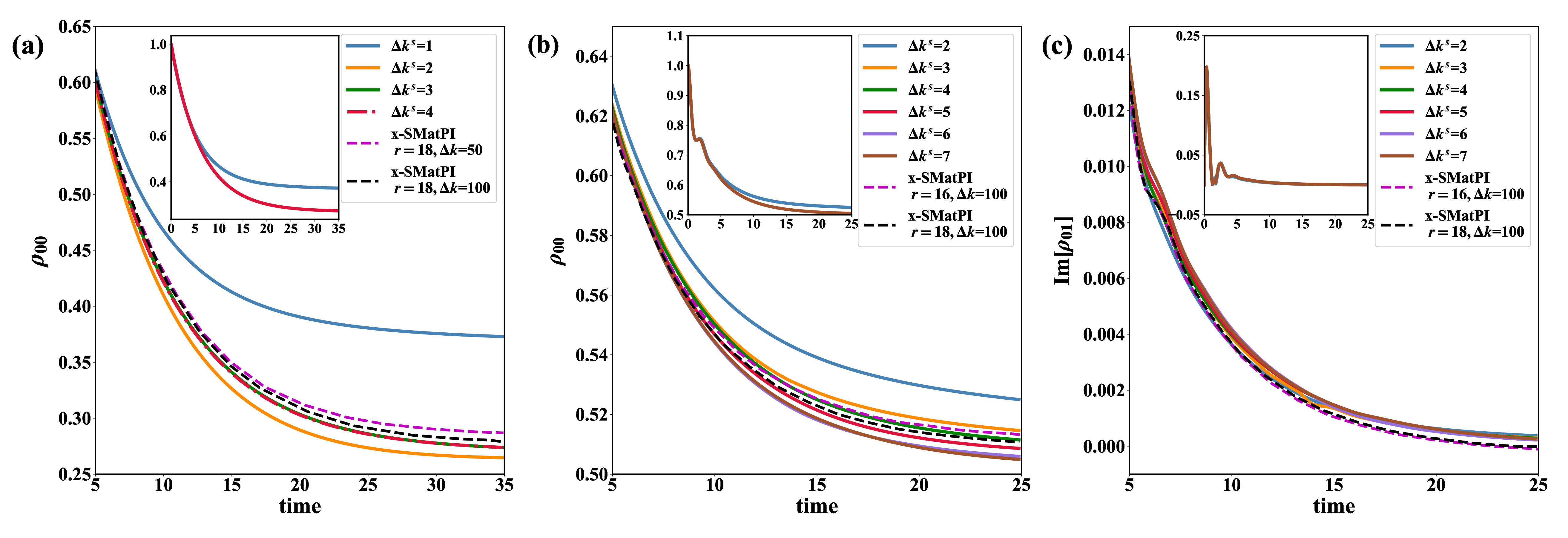}
    \caption{
        (a) Population $\rho_{00}(t)$ for SBM Case \RNum{3} and (b)(c) $\rho_{00}
        (t) $ and $\textrm{Im} \rho_{01}$ for SBM Case \RNum{4} calculated by
        ST-QuAPI with SVD cutoff $\xi=10^{-6}$ and with different $\Delta k$.
        The dashed lines are the result calculated by x-SMatPI in
        Ref.~\onlinecite{makri2021small}. The detailed parameters of SBM and
        other setups are given in the text. 
        }
    \label{fig:fig4}
\end{figure*}

\subsection{SBM with Ohmic spectral density}
The next two cases we select are what have been studied in detail by Makri in
Ref.~\onlinecite{makri2021small} with the newly developed small matrix path
integral with extended memory approach (x-SMatPI). It has been shown that the
results calculated by the conventional iQuAPI approach ($\Delta k=18$) are far
from the converged results. We use ST-QuAPI to simulate the dynamics of the two
cases and compare the results with that of x-SMatPI.  Ohmic spectral density is
adopted in these two case, $J(\omega)=\frac{1} {2}\pi \alpha \omega
e^{-\omega/\omega_c}$. The parameters are $\varepsilon=5$, $\Delta=-1$,
$\omega_c=2$, $\alpha=4$, $\beta=0.1$ in Case \RNum{3} and $\varepsilon=0$,
$\Delta=-1$, $\omega_c=1$, $\alpha=2$, $\beta=1$ in Case \RNum{4}. The initial
bath state is in equilibrium with the up spin state, which is realized by
shifting the coordinate of the system in our
simulation.~\cite{walters2015iterative}
The time step size and cutoff is $\Delta t=0.03$, $\xi=10^{-6}$ for Case \RNum{3}
and $\Delta t=0.125$, $\xi=10^{-6}$ for Case \RNum{4}.

Fig.~\ref{fig:fig4}(a) shows the results for Case \RNum{3}. The population
$\rho_{00}(t)$ of ST-QuAPI with $\Delta k=3$ has already converged within
$t<35$. The black and violet dashed curve is calculated by x-SMatPI with $r=18,
\Delta k=100$ and $r=18, \Delta k =50$. In SMatPI, $r$ is called the
entanglement length, the detailed definition of which can be found in the original
paper.~\cite{makri2021small} The larger the value of $r$ the more accurate the
result.  The time step size used in x-SMatPI is $\Delta t=0.0625$ different from
ST-QuAPI. The maximum discrepancy between the result of ST-QuAPI and the most
accurate result of x-SMatPI ($r=18, \Delta k=100$) is less than 0.01. The number
of renormalized states of ST-QuAPI is 4, 12, 17, 21 with $\Delta k=1, 2, 3, 4$,
respectively.  Case \RNum{4} is more demanding than Case \RNum{3}. Fig.~\ref{fig:fig4}(b)(c) 
shows $\rho_{00}(t)$ and $\textrm{Im} \rho_{01}(t)$ for Case \RNum{4}
. The discrepancy between ST-QuAPI with $\Delta k=7$ and x-SMatPI with $r=18,
\Delta k=100, \Delta t=0.25$ is also smaller than 0.01. The number of
renormalized states of ST-QuAPI is 12, 28, 36, 43, 52, 54 with $\Delta k$ from 2
to 7, respectively.  The comparison between these two approaches not only
demonstrates the correctness of the results, but also reveals the effectiveness
of ST-QuAPI in capturing long memory effect by influence functional
renormalization.

\section{Conclusion and outlook}
\label{sec:conclusion}

In this work, we propose a new memory truncation scheme for iterative QuAPI to
simulate the quantum dynamics of system-bath coupled problems. The conventional
memory truncation scheme used in iterative QuAPI discards all the influence
functional beyond a preset time interval, which is not
effective for problems with long memory time. Instead, our memory truncation
scheme selects and retains the most important parts of the originally discarded
influence functional to improve the accuracy. The criterion for selection is to
minimize the difference between the augmented reduced density tensor before and
after truncation, realized by the density matrix renormalization group
algorithm. As a result, an auxiliary time slice that contains partial
information of the history is constructed adaptively and iteratively, and so is
the influence functional between the effective time slice and future time
slices. Therefore, the new memory truncation scheme is more effective than the
conventional memory truncation scheme, especially for problems with long memory
time. This is the main contribution of this work.

We have demonstrated the effectiveness of the improved memory truncation scheme by
simulating the quantum dynamics of the spin-boson model. Four different
parameter sets are adopted, including Debye spectral density and Ohmic spectral
density. In all four cases, our scheme converges much more quickly with the
truncation length than the conventional memory truncation scheme. By examining
the singular values of ARDT over time, whose distribution characterizes the
temporal correlation of the system, we found that in the studied cases, the
singular values decay quickly with the indices of renormalized states and the
total temporal correlation over time is bounded. This observation indicates the
effectiveness of selecting basis states via the DMRG algorithm.

This new memory truncation scheme is also compatible with the recently developed
time-evolving matrix product operator approach. It uses a matrix product state
to approximate the exponentially large augmented reduced density tensor, thereby
greatly reducing the computational cost, especially suitable for problems with
multiple system states and long memory time. The same truncation scheme as
iQuAPI has been adopted to simulate long-time dynamics. We expect that with our
truncation scheme, the accuracy of TEMPO will further increase. Related studies
are currently being carried out in our group.

\section*{SUPPLEMENTARY MATERIAL}
See the supplementary material for the proof of exactness of ST-QuAPI when
$\xi=0$ and the change of the number of renormalized states and singular value
distribution with time of SBM Case \RNum{1} as Fig.~\ref{fig:fig3}.

\begin{acknowledgments}
    The authors are grateful to Professor Jiushu Shao for the insightful
    discussion. This work is supported by the Innovation Program for Quantum
    Science and Technology (Grant No. 2023ZD0300200), the National Natural
    Science Foundation of China (Grant No. 22273005), NSAF (Grant No. U2330201),
    and the Fundamental Research Funds for the Central Universities.    
\end{acknowledgments}

\section*{Conflict of interest}
The authors have no conflicts to disclose.

\section*{Data Availability}
The data that support the findings of this study are available from the corresponding author upon reasonable request.

\bibliography{reference}% Produces the bibliography via BibTeX.

\end{document}